\documentclass[letter]{aa} 

\usepackage{graphicx}
\usepackage{txfonts}
\usepackage{microtype}
\begin{document}

   \title{Detection of intended and unintended emissions from Starlink satellites in the SKA-Low frequency range, at the SKA-Low site, with an SKA-Low station analog}

   \titlerunning{Detection of Starlink satellites at low frequencies}

   \author{D. Grigg
          \inst{1,2},
          S. J. Tingay\inst{1},
          M. Sokolowski\inst{1},
          R. B. Wayth\inst{1},
          B. Indermuehle\inst{3}, and
          S. Prabu\inst{1}
          }

    \authorrunning{Grigg et al.}

   \institute{International Centre for Radio Astronomy Research, Curtin University, Bentley, WA, 6102, Australia\\
              \email{s.tingay@curtin.edu.au}
         \and
             DUG Technology, 76 Kings Park Rd, West Perth, 6005, WA, Australia
        \and
            CSIRO Space \& Astronomy , PO Box 76, Epping, NSW, 1710, Australia
             }

   \date{Received 2023-08-04; accepted 2023-09-27}


   \abstract
   {Intended and unintended radio emissions from satellites can interfere with sensitive radio telescopes in the frequency ranges of key experiments in astrophysics and cosmology.  We detect strong intended and unintended electromagnetic radiation from Starlink satellites at the site of the future SKA-Low facility in Western Australia, using an SKA-low prototype station known as the Engineering Development Array version 2 (EDA2).}
   {We aim to show that Starlink satellites are easily detectable utilising a configuration of low frequency radio antennas representative of an SKA-Low ``station'' and that our results complement similar findings with the LOFAR telescope.}
   {Utilising the EDA2 at frequencies of 137.5 MHz and 159.4 MHz, we detect trains of Starlink satellites on 2023-03-17/18 and 2021-11-16/17, respectively, via the formation of all-sky images with a frequency resolution of 0.926 MHz and a time resolution of 2 s.  Time differencing techniques are utilised to isolate and characterise the transmissions from Starlink and other satellites.}
   {We observe Starlink satellites reaching intensities of $10^6$ Jy/beam, with the detected transmissions exhibiting a range of behaviours, from periodic bursts to steady transmission.  The results are notable because they demonstrate that Starlink satellites are detected in the SKA-Low frequency range, transmitting both intentionally and unintentionally. Follow-up work and discussion are needed to identify the cause of this unintentional radiation as it has the potential to interfere with SKA-Low science. It is likely that the transmission levels will need to be reduced by orders of magnitude to bring the impact on radio astronomy to potentially manageable levels.}
   {Our results indicate that both intended and unintended radiation from Starlink satellites will be detrimental to key SKA science goals without mitigation. Continued conversation with SpaceX could potentially result in future mitigations which the EDA2 instrument could efficiently monitor and characterise at the SKA-Low site.}

   \keywords{Radio astronomy, space situational awareness, SKA-low, Starlink, unintended electromagnetic radiation}

   \maketitle
%

\section{Introduction}
As the most sensitive radio telescopes ever conceived are under construction, satellite constellations are being deployed in low and medium Earth orbit (LEO/MEO) at an unprecedented scale, in support of global communications, navigation, and remote sensing (among other applications)\footnote{\url{https://planet4589.org/space/con/conlist.html}}.  

The intended and unintended electromagnetic radiation (IEMR and UEMR) from satellites in LEO and MEO now form a major source of radio frequency interference (RFI) for radio telescopes on Earth. This includes those that are situated in previously pristine locations, many of them protected by Radio Quiet Zones (RQZs) via their respective national regulators.  By design, satellite constellations are visible from any location on Earth (depending on constellation parameters such as inclination).  Thus, no location can be protected from electromagnetic radiation emitted from satellites unless adequate regulatory means are established. Until then, astronomers depend on the engagement of satellite operators in good faith with our community to mitigate these impacts. 

In this work, UEMR refers to unintentional electromagnetic radiation not originating from a satellite's transmitter system, consistent with terminology used in recent work by \citet{lofar}. Examples of UEMR could include radiation from navigation computers, high speed network switches, or Hall effect thrusters, power supplies, or any other type of electronic equipment on board a satellite. Some of these sources of UEMR may be weak, but from a radio astronomy point of view, they can have a serious impact on science.

IEMR is referred to as intentional electromagnetic radiation originating from a satellite's transmitter system which has been approved by relevant governing authorities. Satellites are legally able to transmit at designated frequencies that are allocated for this purpose after approval is given via international governing bodies such as the radiocommunications sector of the International Telecommunication Union's (ITU) Radio Regulations (ITU-R)\footnote{\url{https://www.itu.int/pub/R-HDB-22-2013}}. This can therefore be referred to as `emission' as per the ITU-R definition.

The specific case considered in this Letter is related to low frequency radio astronomy and the detection of IEMR and UEMR in this frequency range from satellites in the Starlink constellation.  We have focused on the Starlink constellation because, at the current time, they are the largest constellation with more than 4,000 satellites in orbit. In particular, we consider the low frequency component of the Square Kilometre Array \citep[SKA,][]{2009IEEEP..97.1482D}, currently under construction at Inyarrimanha Ilgari Bundara, the CSIRO Murchison Radio-astronomy Observatory in Western Australia.  

Within the SKA-Low frequency range (50 - 350 MHz), the only frequency bands that are protected by the ITU-R for radio astronomy are as follows: 73 - 74.6 MHz, 150.05 - 153 MHz, 225 - 235 MHz, and 322 - 328.6 MHz. It is important to note that from an international and national regulatory point of view, only transmissions received by radio telescopes in these frequency ranges in the designated regions are considered interference. 

Despite this, radio astronomy has always depended on being able to observe the Universe at all wavelengths. In the case of SKA-Low, the frequency range of 100 - 200 MHz is key to the SKA headline science case related to detecting emissions from redshifted neutral hydrogen from the Epoch of Reionisation \citep[EoR, the period up to a billion years after the Big Bang, when the first stars and galaxies formed,][]{eor}.

Hosted at the SKA-Low site, the Engineering Development Array version 2 \citep[EDA2,][]{eda2} is an array of 256 low frequency bowtie antennas, arranged in the configuration of an SKA-Low station (35 m diameter physical footprint). The EDA2 has been used as a ground-truth system for the evaluation of SKA-Low stations \citep{2020IOJAP...1..253B} and can be operated in an all-sky imaging mode.  The EDA2 has undertaken RFI surveys at the SKA-Low site in the FM band, in the process detecting both reflected FM emissions and both IEMR and UEMR from objects in LEO \citep{tingay_eda_ssa,grigg}.  Using the EDA2 at other frequencies (159.4 and 229.7 MHz), in a search for astronomical transients, \citet{soko_eda2} also detected UEMR from satellites.

Of specific relevance is a recent publication describing observations of Starlink satellites by the Low Frequency Array \citep[LOFAR,][]{lofar_desc}, across 110 - 188 MHz. Broadband radiation was detected that occupies the entire frequency range, as well as narrow-band emissions at 125, 135, 143.05, 150, and 175 MHz (emissions at 143.05 MHz are attributed to reflections of the GRAVES radar system transmitter) \citep{lofar}.  The measured flux densities range between 0.1 and 10 Jy for the broadband radiation and between 10 and 500 Jy for the narrow-band radiation.  The authors provide a comprehensive description of their observations, specifically in the context of the regulatory environment for UEMR from objects in space, as well as a discussion of the likely consequences for radio astronomy if this radiation cannot be mitigated.

The intent of this Letter is to present relevant data from the EDA2, arising from our continued work to characterise the RFI environment at the SKA-Low site \citep{tingay_eda_ssa,grigg}.  We briefly present data and analysis pertaining to the detection of Starlink satellites with the EDA2, complementing results from the more comprehensive analysis done with the LOFAR telescope.  Our data serve to confirm that radiation from Starlink satellites is detectable from the SKA-Low site, in the SKA-Low frequency range, and with an instrument representative of the sensitivity and general characteristics of an SKA-Low station. Consequently, we are able to qualitatively characterise potential impacts to observations made with the SKA-Low and show that RQZs operating at a national level are insufficient for the regulation of radiation originating from satellites.  We also demonstrate a novel ``all-sky'' approach to the detection and characterisation of the radiation of interest, which is different to the approach of the LOFAR work.

Section \ref{sec:methods} briefly describes observational parameters and data processing methods.  In Section \ref{sec:results}, the results of those observations and data processing are described, followed by a discussion and conclusions in Section \ref{sec:discussion}.

\section{Observations and data processing}
\label{sec:methods}
Two sets of observations were made with the EDA2.  The first set took place between UTC 2023-03-17 12:19:09 and UTC 2023-03-18 07:56:05, a duration of 19.5 hours, at a central frequency of 137.5 MHz. The second set of observations took place between UTC 2021-11-16 01:24:18 and UTC 2021-11-17 00:37:07, a duration of approximately 23 hours, at a central frequency of 159.4 MHz.  In both cases, the bandwidth was 0.926 MHz and both orthogonal linear polarisations (X - east-west and Y - north-south) of the EDA2 were utilised. 

These frequencies were chosen as part of a larger body of work, where 137.5 MHz is a known ORBCOMM satellite downlink frequency and 159.4 MHz is a well established test frequency for the EDA2 due to minimal RFI and an optimal response for the MWA dipole and signal chain.

The voltage data generated by the EDA2 system were used to form cross-correlation products between all independent pairs of the 256 antennas (forming XX, YY, XY, and YX correlation products).  The correlator output was averaged over 2 seconds and the full 0.926 MHz band, forming visibility data at those resolutions, which were calibrated in amplitude and phase as described by \citet{soko_eda2}.

Full sky images using the parallel hand polarisation visibilities, XX and YY, were produced by processing the visibility data with the \texttt{MIRIAD} package \citep{miriad}. For more detail on this procedure, refer to \citet{tingay_eda_ssa}. For the 137.5 MHz (159.4 MHz) dataset, the images were 160 $\times$ 160 (180 $\times$ 180) pixels in size, each pixel was 1.2$^{\circ}$ (1.0$^{\circ}$) in extent with an approximate angular resolution of 3.6$^{\circ}$ (3.1$^{\circ}$), and 71,252 (83,426) images were made, which is an aggregate across both polarisations.

\begin{figure*}[h]
    \centering
    \includegraphics[width=1\textwidth]{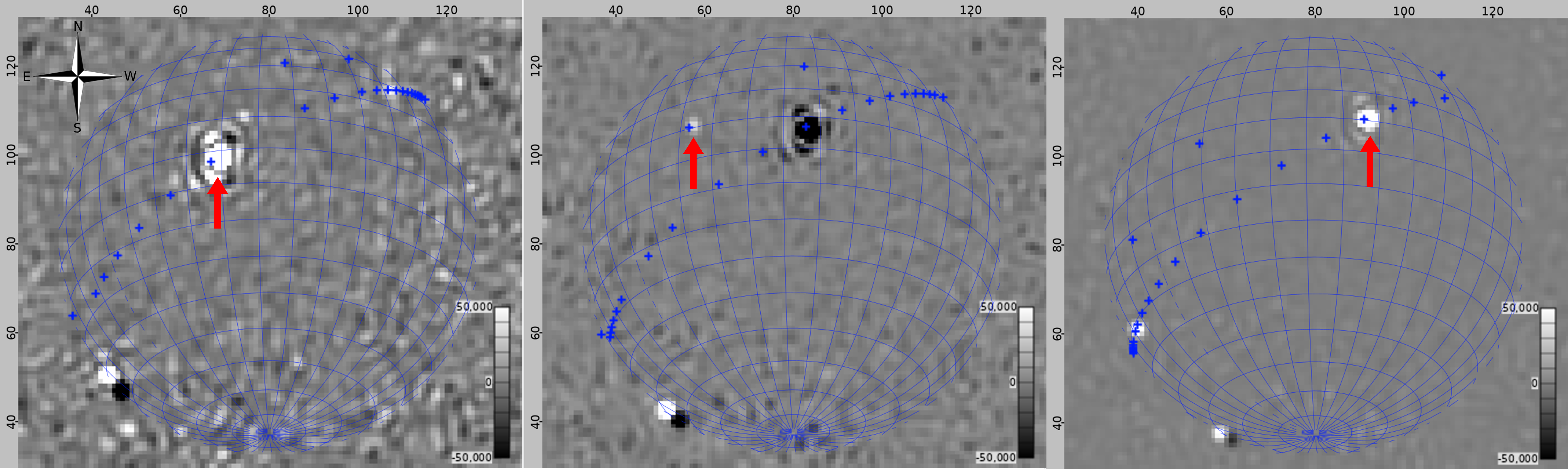}
    \caption{\label{fig:images_137} Three images of Starlink satellites detected at 137.5 MHz. The images from left to right are at times UTC 2023-03-18 00:59:46, 01:01:04, and 01:02:24. The blue grid is an astronomical projection to show the area of visible sky. The blue crosses are the TLE predicted positions of all Starlink satellites detected during this dataset. The red arrows denote detections of NORAD 55,358, NORAD 55,377, and NORAD 55,340 from left to right. The satellite in the bottom left corner of all three images is not a Starlink satellite (NORAD 28,654 - NOAA 18).  Arrows indicate the detections referred to. The median of the Median Absolute Deviation (MAD) value at the zenith in each image across all difference images at this frequency is 310 Jy/beam and is high due to there nearly always being a satellite transmitting overhead throughout the entire dataset. The MAD of images was as low as $\sim$0.6 Jy/beam when there were no transmitters or Milky Way or Sun present.}
\end{figure*}

The images were input to a Python-based detection script. Following our previous work \citep{grigg}, the images for the two polarisations were averaged and differenced in time. The differencing of images closely separated in time removes the majority of astronomical signals, while isolating detections of faster moving objects in orbit.  In this case, we difference images separated by 16 time steps (32 seconds), such that $\Delta I_{n}=I_{n}-I_{n - 16}$, where $I$ is the image at image index $n$.  Satellite detections appear in the difference images with positive amplitudes if present in the $I_{n}$ image and with negative amplitudes if present in the $I_{n - 16}$ image. Astronomical sources do move across the sky enough to produce weak residuals in the difference images using a time separation of 32 s, but this timescale ensures that Starlink satellites close to the horizon (moving with a lower angular speed) travel far enough to appear in the difference images. Examples of this technique can be found in \citet{steve2, tingay_eda_ssa, grigg}. 

In order to identify any orbiting objects in our difference images, Two Line Elements (TLEs) were queried from space-track.org for the entire public catalogue, over the period 2023-03-16/20 for the 137.5 MHz observations and over the period 2021-11-15/19 for the 159.4 MHz observations.  The TLEs were then used to predict the positions of all objects in each image of the datasets.

The results were inspected and analysed using the software tool DUG Insight \citep{grigg}, which provided an interface to efficiently interrogate the data. Figures \ref{fig:images_137} and \ref{fig:images_159} were produced using this software.

\section{Results}
\label{sec:results}

A comparison of the predicted positions of all satellites from publicly available TLEs with the emissions detected in the difference images revealed that Starlink satellites had been detected at both frequencies.

\subsection{137.5 MHz}
\label{137_results}
An example sequence of difference images from the 137.5 MHz data are shown in Figure \ref{fig:images_137}. In the difference images, it is apparent that the emission from these Starlink satellites is highly variable. This is more readily apparent in a video of the 137.5 MHz difference images, showing the behaviour of the emission from the Starlink trains\footnote{\url{https://youtu.be/TT1hJ2NOZQo}}.

The detected emissions are a series of flashes from each satellite.  The frequency of the flashing is approximately once per $\sim100$ seconds (every 50 or 51 images).  Three satellites are exceptions. NORADs 55,572, 55,582, and 55,612 flash at a slightly longer period of 120 seconds.  In total, over the observation period, 1,130 flashes were detected from 136 Starlink satellites in the NORAD \# range 53,436 to 55,715.

These detections were made by force-fitting a Gaussian model at the predicted position of each satellite in each polarisation, and accepting fits where the percentage errors on the amplitude and width of the Gaussian are $<$10\%.  No detections were attempted at elevations below 5$^{\circ}$ or within 2$^{\circ}$ of the Sun.  Where the detection is consistent with two or more Starlink satellites, the closest on the sky is taken as the identification.

Emission intensities were estimated by correcting the XX and YY measured intensities using the antenna beam response models \citep{mwa_beam} and summing the two polarisations together.  The flashes reach a maximum intensity of approximately $10^{6}$ Jy/beam at ranges of $\sim$500 km (EIRP $\sim$ 30 mW) and a minimum intensity of approximately 2,000 Jy/beam at ranges of $\sim$2,000 km (EIRP $\sim$ 1mW).  EIRPs are calculated assuming the emission occupies the full 0.926 MHz bandwidth of the system. It is important to note that these emissions are highly localised and constitute only a minor portion of the total power entering the EDA2 system meaning that the response of the EDA2 will still be linear. Multiple passes were observed for the majority of these objects.  The flashes are less than 2 seconds in duration (since they occupy at most two consecutive images).  Thus, pairs of images were averaged together to ensure 100\% of the emission was captured, making the integration time 4 seconds for the measurement of the flashes.

\begin{figure}
    \centering
    \includegraphics[width=0.5\textwidth]{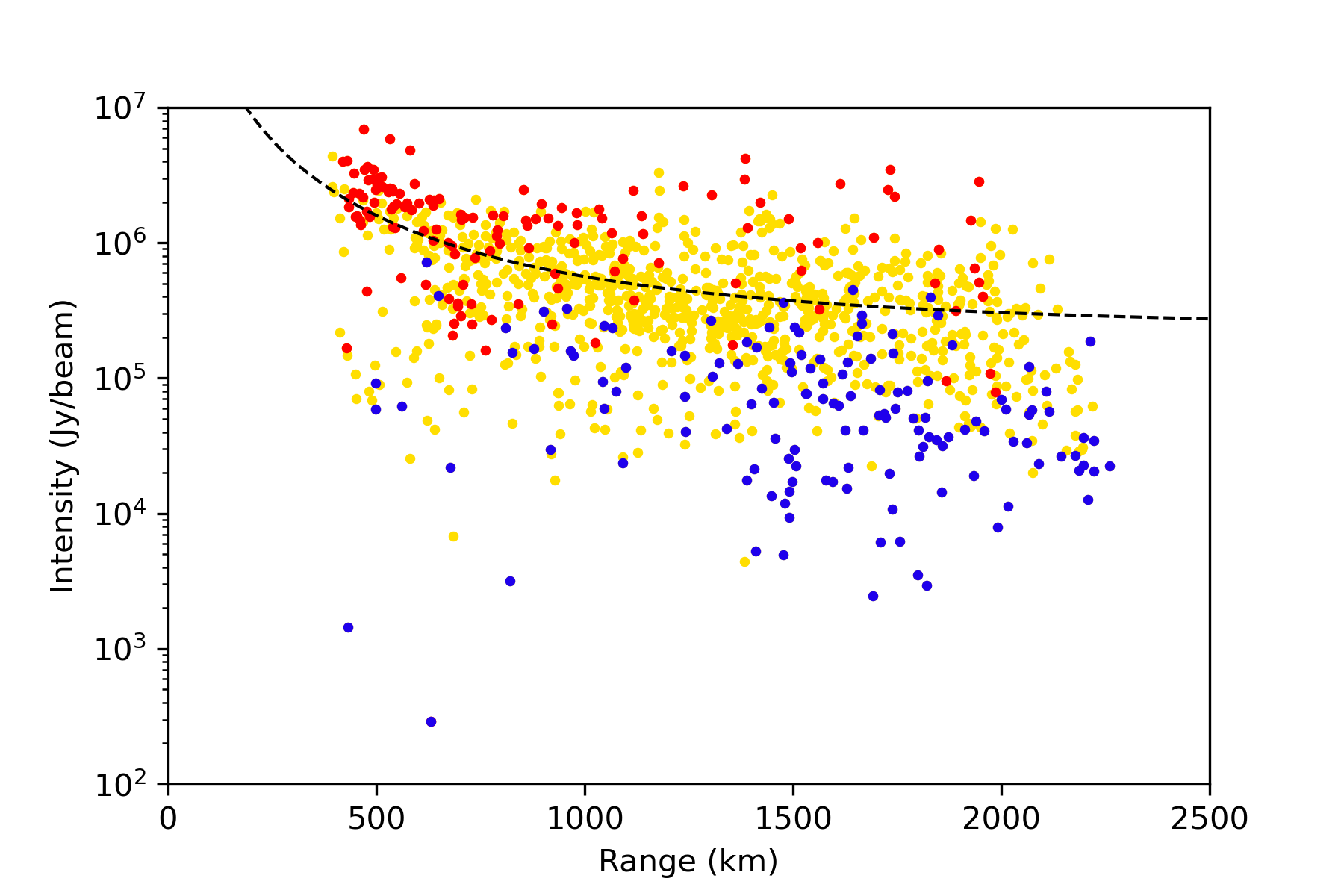}
    \caption{\label{fig:flashes_137} The range of peak intensity measurements for all 1,130 flashes detected at 137.5 MHz. The red points represent the flash with the highest intensity for each NORAD identifier, while the blue points represent the lowest. The yellow points represent the rest of the flashes. An inverse square law ($\propto \frac{1}{r^2}$) is fitted to these data points to account for the spherical isotropic propagation of the transmission and to guide the reader's eye.}
\end{figure}

Figure \ref{fig:flashes_137} shows the relationship between peak intensity and range ($r$) for all flashes detected in the dataset.  In general, weaker emissions occur at larger ranges, but intrinsic variability in the emission strength is also apparent for individual satellites, as well as between satellites. 

\begin{figure*}[h]
    \centering
    \includegraphics[width=1\textwidth]{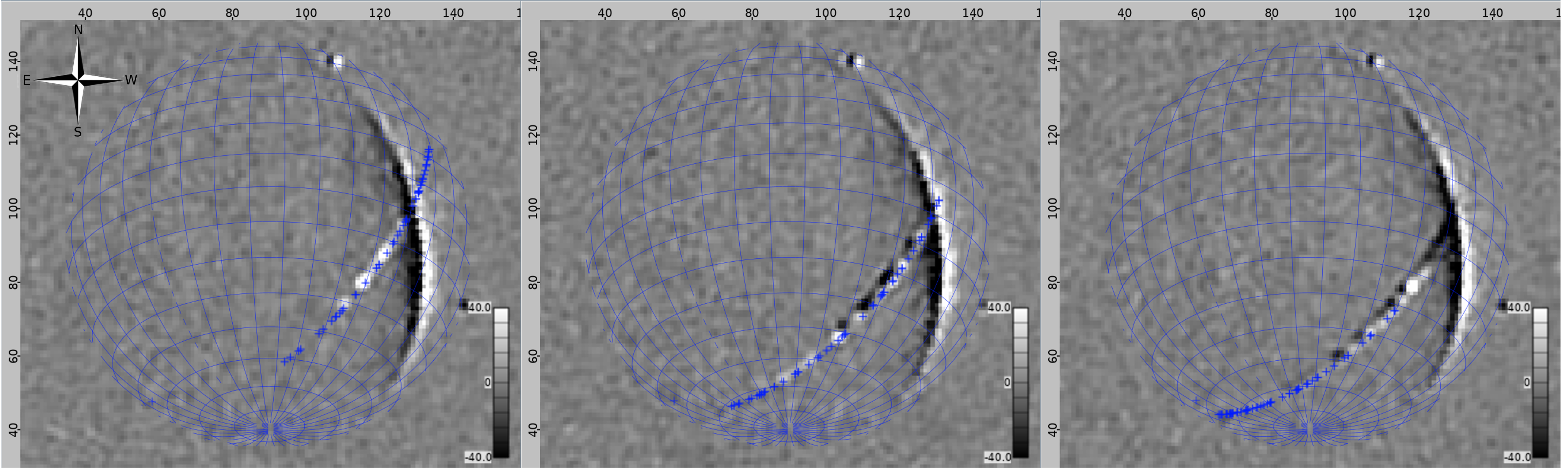}
    \caption{\label{fig:images_159} Three images of the detected Starlink train at 159.4 MHz. The images from left to right are at times UTC 2021-11-16 10:07:10, 10:07:49, and 10:08:29. The blue grid is an astronomical projection to show the area of visible sky. The blue crosses are the TLE predicted positions of all Starlink satellites detected during this dataset. The vertical band on the right hand side of the image is due to the motion of the Milky Way over the differencing period. The median MAD value at the zenith across all difference images is 1.2 Jy/beam.}
\end{figure*}

The 137.5 MHz frequency overlaps with a standard satellite communications downlink frequency band, meaning that many other satellites were detected in this dataset. Very high intensity emissions from constellations such as SPACEBEE, ORBCOMM, and NOAA reached 10$^{7}$ Jy/beam (highest recorded near the zenith was NOAA 18 at 6.6$\times10^{7}$Jy/beam), causing extreme sidelobes in the images which obscured weaker emissions. 

SpaceX engineers have confirmed that the Starlink satellites have `Swarm' transmitters, which transmit at 137 MHz (2023, SpaceX, private communication). These transmitters are used for telemetry, tracking and control, which could explain why they are observed as sparse, periodic emissions. This frequency band will continue to be used by Starlink satellites for the foreseeable future.  These emissions are in the class of IEMR.

A report by the European Communications Committee\footnote{https://docdb.cept.org/download/4013} suggests that the Swarm transmitters have an EIRP downlink limit of -1.55 dBW ($\sim$700 mW) over a 42 - 125 kHz band.  Taking our observed 30 mW EIRP near minimum range and correcting for the bandwidth and our upper limit on the emission duration, we get a lower limit of $\sim$200 mW (125 kHz transmit band and $<$2 s duration), consistent with the published value.  The document also suggests a 0 dBi transmit antenna, which appears inconsistent with the scatter in received power in Figure \ref{fig:flashes_137}. This can be primarily attributed to variable transmission power of the transmitters which can be seen in the vertical spread of points at each range.

\subsection{159.4 MHz}
\label{159_results}
An example sequence of difference images from the 159.4 MHz data are shown in Figure \ref{fig:images_159}. The radiation from satellites in the Starlink train is constant, rather than the flashing observed at 137.5 MHz. A video of the 159.4 MHz data shows the behaviour of the radiation from this train\footnote{\url{https://youtu.be/FISUgjrCAi4}}.

\begin{figure}
    \centering
    \includegraphics[width=0.5\textwidth]{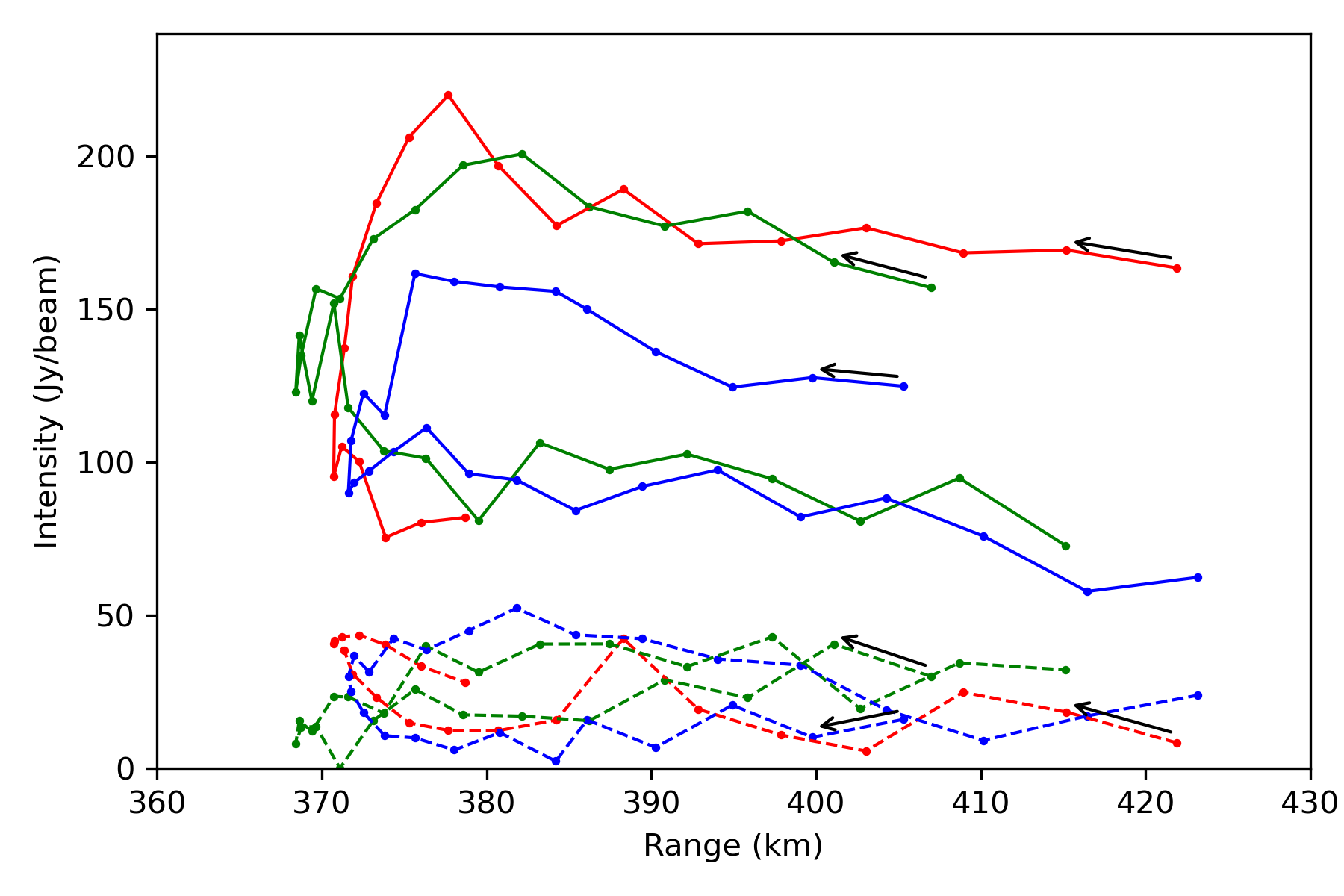}
    \caption{\label{fig:amps_159} The range of intensity measurements of 3 satellites fitted in the train of Starlink satellites at 159.4 MHz. The three colours identify the particular satellite fitted. The dotted lines are for the XX polarisation images and the solid lines are for the YY polarisation images. The arrows show the direction of time.}
\end{figure}

At 159.4 MHz, the steady radiation observed reaches a peak intensity of $\sim$220 Jy/beam (at a distance of $\sim$380 km, corresponding to an approximate EIRP of $\sim$2$\mu$W). Again, the measured intensities are corrected by the beam models.  The identification of individual satellites is challenging, due to the small angular separation of the satellites relative to the angular resolution of the EDA2.  However, predicted positions generated from TLEs suggest that this train includes 53 Starlink satellites with NORAD IDs between 49,408 and 49,460.  This is a different set of satellites compared to those detected at 137.5 MHz.

These satellites were launched on 2021-11-13, on a FALCON 9 rocket. The observations at 159.4 MHz were taken only 3 days after this launch, allowing an opportunity to study the Starlink satellites at low altitude. The nominal smallest range to this Starlink train was $\sim$370 km, as predicted by the TLEs. They later ascended to higher orbital altitudes at 500 - 600 km.

The predicted positions lag behind the observed radiation in the dataset, which is likely due to the time difference between the epoch of the TLE used and the time of the detection. The first available TLE for this train was generated the day after this observation, at UTC 2021-11-17 16:00. This is a difference of 21 hours. As the Starlink satellites would have been rising to their operational altitude at this time, it is reasonable that we see a mismatch between the detected and predicted positions. The same Starlink train was not detectable in four other passes in the period of the dataset. These were at significantly lower elevation angles with two being at a range of $\sim$750 km and the other two at $\sim$1,750 km. Assuming the same power of the radiation for these other passes, taking into account range and primary beam attenuation it is not expected that a detection would be made at these ranges.

Figure \ref{fig:amps_159} shows how the intensity of three satellites in the train varies as a function of range for each polarisation. As the intensities of satellites in the train were only high enough to fit a Gaussian model in the YY polarisation images, the fitting procedure was carried out on just YY polarisation images, allowing for the determination of the peak location. Subsequently, the intensity value at this specific pixel position was extracted from both polarisations and are the values reported in Figure \ref{fig:amps_159}. The ranges were calculated by identifying the Starlink satellite closest to the fit position.  Although there are some uncertainties in the identifications, the ranges are suitable approximations.

Figure \ref{fig:amps_159} shows a distinctive evolution of the radiation intensities in the YY polarisation, starting high and going through a transition to much lower radiation levels at closest approach to the EDA2.  The XX polarisation intensities mirror that behaviour in reverse (start low and transition high), but at lower intensity levels.  This may suggest that geometric effects due to the transmission surface being obscured at particular orientations as the satellite passes over could be causing this behaviour.

The authors have been in communication with SpaceX (who owns, builds, and operates the Starlink constellation), who explain that this radiation is likely due to the satellites' propulsion or avionics system and is likely over 50 - 200 MHz (2023, SpaceX, private communication). The propulsion system is actively engaged during the time this train is detected. This radiation is therefore in the class of UEMR.

During the period of observation, there were 114 additional Starlink satellites accessible to our data, of the same generation as the Starlink satellites we have detected. There are many passes of these objects, with a closest range of $<$450 km.  None of these objects display similar UEMR to the detected train.



\section{Discussion and conclusions}
\label{sec:discussion}

The received power of transmissions from the detected Starlink satellites was often much higher than the brightest astronomical radio sources in the sky. While the detected IEMR at 137.5 MHz complies with ITU-R regulations, it is important to note that the UEMR detected at 159.4 MHz poses interference problems for low frequency radio astronomy. It is important to reiterate that the frequencies utilised for these observations are not within the ITU-R protected bands for radio astronomy, but are present in bands that are relevant to the highest impact astrophysics planned for the SKA-Low facility.

These observations at 137.5 and 159.4 MHz are not simultaneous, so we can draw no conclusions regarding the frequency extent of the transmissions we detect.  Although the frequency extents of this work are different to the LOFAR work, both studies detect Starlink satellites across a large fraction of the SKA-Low frequency range \citep{lofar}. 

It is of interest to note that the \citet{lofar} work did not detect the Swarm transponder IEMR, which may be due to geographical restrictions such as not transmitting over Europe. Also, satellites in the \citet{lofar} paper which were still being boosted to their operational altitude were detected at flux densities up to 10 Jy. Not mentioned by the authors, multiple generations of Starlink satellites were detected and could explain the variation in fluxes reported in that work. This makes directly comparing fluxes between the \citet{lofar} work and this Letter complex, as well as comparing between instruments, frequencies, timeframes and geographical locations. This being said, the $\sim$0.1-0.3 Jy signals detected at 550 km range in the \citet{lofar} work would not be expected to be detected in this work as this below the threshold of even images with the lowest MAD values for each dataset. For a comprehensive cross-examination between the two studies, further observations and perhaps collaboration with SpaceX is needed to account for these differences.

Broad and narrow-band IEMR and UEMR could severely compromise key science for SKA-Low, such as measurements of the EoR \citep{Furlanetto_2006}. \citet{2020MNRAS.498..265W} estimated the impact of RFI on the measurement of the EoR power spectrum (the observational quantity used to detect or constrain the EoR signal) and determined that the presence of an aggregate apparent 1 mJy of RFI in the integrated power spectrum is enough to prevent detection for even the most optimistic theoretical models of the EoR signal. After all forms of avoidance and RFI mitigation and removal, including attenuation due to averaging in frequency and time, the apparent RFI remaining in the data needs to be below $\sim$mJy levels.


An engineering solution to mitigate the IEMR and UEMR at the source on the satellites may be an option. In this case, reductions in the EIRP of factors of order 100 to 1,000 may be required to bring the radiation into a regime that does not challenge key radio astronomy experiments (assuming a notional temporal occupancy of 10\% for 100 Jy radiation over a 1 MHz band within an overall 100 MHz observation bandwidth). A collaborative effort between astronomers and satellite operators could explore engineering solutions.

The authors are in close contact with SpaceX, and the company has offered to continue to discuss possible ways to mitigate any adverse effects to astronomy in good faith. As part of their design iteration, SpaceX has already introduced changes to its next generation of satellites which could mitigate the impact of the Swarm transponder emissions. This highlights the importance of studies such as this and \citet{lofar} at identifying UEMR, so that policies can be developed based on quantitative information, to ensure the private LEO space industry can operate whilst minimising its impact on radio astronomy research.

If engineering solutions cannot be found, or cannot bring the radiation under an appropriate threshold, RFI identification and flagging algorithms could be applied to the data collected from radio telescopes.  Substantial effort has been applied in this direction, for example as described in \citet{offringa2015} and \citet{Wilensky_2019}, and in a particularly relevant study, by \citet{2023MNRAS.524.3231F}. Flagging of data comes as a last resort, as valuable information is also lost.


SpaceX's engagement with the astronomy community has allowed us to confirm that the measured UEMR is originating from their current generation satellite platforms. In the absence of suitable regulation, we thus depend on satellite operators engaging with the radio astronomy community to find solutions. We encourage all satellite operators to seek our engagement before launching their satellites, to ensure adequate UEMR mitigation is in place on their platforms. The appropriate venue to establish this contact is via ITU-R Working Party 7D (Radio Astronomy), where radio astronomers meet several times per year to protect the radio spectrum and assess other users' needs.

In the domain of optical astronomy, significant progress has been made in mitigating the reflected sunlight from Starlink satellites that interferes with optical telescope observations.  \citet{2023arXiv230606657M} reports that Starlink Generation 2 satellites are a factor of 12 fainter than Generation 1 satellites, despite the larger size of Generation 2 satellites.  This demonstrates that mitigations are possible.

The EDA2 has proven to be a very effective and efficient tool for measuring and characterising UEMR from objects in LEO, due to its sensitivity, all-sky imaging capability, and its radio quiet location.  The EDA2 will be a very effective tool for monitoring UEMR from all satellites in the SKA-Low frequency range, including to assess any improvements realised from mitigations.

In summary, our conclusions are:

   \begin{enumerate}
      \item Using observations made with the EDA2 at 137.5 and 159.4 MHz on two different occasions, we detect both IEMR (telemetry beacons) and UEMR (propulsion or avionics; 2023, SpaceX, private communication) respectively from trains of Starlink satellites.
      \item Intensities of IEMR at 137.5 MHz are of order 10 $-$ 10$^{6}$ Jy/beam, or in the $\mu$W - mW EIRP range. Although this IEMR abides by ITU-R guidelines, these intensities are large compared to the strongest astronomical radio sources in the sky and will therefore have the potential to disrupt astronomical observations at SKA-Low frequencies;
      \item The detected IEMR and UEMR are outside of the frequency bands protected for radio astronomy, but are at frequencies of great interest for key experiments for the SKA-Low facility, and at frequencies where RQZ protections at the SKA-Low site are in place;
      \item Our results complement recent results published from observations with the LOFAR telescope.  Our results confirm that the SKA-Low site is impacted in the frequency range of the SKA-Low telescope;
      \item Mitigations will be required, either to suppress the UEMR at the source or identify and remove affected sections of data. SpaceX has already introduced changes to its next generation of satellites which could mitigate the impact of UEMR; and
      \item The EDA2 will be a valuable tool for ongoing monitoring and assessment of UEMR from satellites in LEO into the future.
   \end{enumerate}

\begin{acknowledgements}
This scientific work uses data obtained from Inyarrimanha Ilgari Bundara / the Murchison Radio-astronomy Observatory. We acknowledge the Wajarri Yamaji People as the Traditional Owners and native title holders of the Observatory site. Support for the operation of the MWA (the host facility for the EDA2) is provided by the Australian Government (NCRIS), under a contract to Curtin University administered by Astronomy Australia Limited. We acknowledge the Pawsey Supercomputing Centre which is supported by the Western Australian and Australian Governments. Inyarrimanha Ilgari Bundara, the CSIRO Murchison Radio-astronomy Observatory and the Pawsey Supercomputing Research Centre are initiatives of the Australian Government, with support from the Government of Western Australia and the Science and Industry Endowment Fund. We also acknowledge time used on DUG Technology's Perth supercomputer and for their usage of DUG Insight.
\end{acknowledgements}

\bibliographystyle{aa} 
\bibliography{starlink}

\end{document}